\documentclass[
superscriptaddress,
aps,
twocolumn,
]{revtex4-1}

\usepackage{amsmath, amssymb}
\usepackage{bbold}
\usepackage{tikz}
\usepackage{empheq}
\usepackage{xcolor}
\usepackage{upgreek}
\usepackage{graphicx}%
\usepackage{bm}
\usepackage{braket}

\begin{document}
	
	\title{Asymmetric Rydberg blockade of giant excitons in Cuprous Oxide}
	
	\author{Julian Heck\"otter}
	\thanks{These two authors contributed equally. }
	\affiliation{%
	  Experimentelle Physik 2,
		Technische Universit\"at Dortmund,
		D-44221 Dortmund, Germany
		}%
	\author{Valentin Walther}%
	\thanks{These two authors contributed equally. }
	\affiliation{Center for Complex Quantum Systems, Department of Physics and Astronomy, Aarhus University, Ny Munkegade 120, DK-8000 Aarhus C, Denmark
	}%
	\affiliation{ITAMP, Harvard-Smithsonian Center for Astrophysics, Cambridge, Massachusetts 02138, USA}
	\author{Stefan Scheel}
	\affiliation{
		Institut f\"ur Physik, Universit\"at Rostock, 
Albert-Einstein-Stra{\ss}e 23-24, D-18059 Rostock, Germany
	}
	\author{Manfred Bayer}
	\affiliation{%
		Experimentelle Physik 2,
		Technische Universit\"at Dortmund,
		D-44221 Dortmund, Germany
	}%
\affiliation{Ioffe Institute,
	Russian Academy of Sciences,
	194021 St. Petersburg, Russia}

	\author{Thomas Pohl}%
	\affiliation{Center for Complex Quantum Systems, Department of Physics and Astronomy, Aarhus University, Ny Munkegade 120, DK-8000 Aarhus C, Denmark
	}%
	\author{Marc A{\ss}mann}
\affiliation{%
	Experimentelle Physik 2,
	Technische Universit\"at Dortmund,
	D-44221 Dortmund, Germany
}%
	
	\email{marc.assmann@tu-dortmund.de}
	
	\date{\today}

\begin{abstract}
The ability to generate and control strong long-range interactions via highly excited electronic states has been the foundation for recent breakthroughs in a host of areas, from atomic and molecular physics \cite{bendkowsky_observation_2009, booth_production_2015} to quantum optics \cite{chang_quantum_2014, Busche2017} and technology \cite{Tiarks2019, Bernien2017,Graham2019}. 
Rydberg excitons provide a promising solid-state realization of such highly excited states, for which record-breaking orbital sizes of up to a micrometer have indeed been observed in cuprous oxide semiconductors \cite{Kazimierczuk2014}.  
Here, we demonstrate the generation and control of strong exciton interactions in this material by optically producing two distinct quantum states of Rydberg excitons. This makes two-color pump-probe experiments possible that allow for a detailed probing of the interactions. Our experiments reveal the emergence of strong spatial correlations and an inter-state Rydberg blockade that extends over remarkably large distances of several micrometers. 
The generated many-body states of semiconductor excitons exhibit universal properties that only depend on the shape of the interaction potential and yield clear evidence for its vastly extended-range and power-law character. 
\end{abstract}

\maketitle

Immersing an exciton into a many-body state of another species gives rise to a number of fascinating phenomena, ranging from the formation of polarons \cite{sidler2017} to the emergence of spinor interactions in semiconducting materials \cite{amo2010, takemura2014}. However, reaching and probing the regime of strong interactions has remained challenging, whereby experimental signatures of interactions are often confined to measurements of spectral line shifts due to short ranged collisional interactions. On the other hand, the strong interactions between Rydberg excitons can act over extraordinary large distances, which suggests new opportunities for probing the fundamental interactions between excitons. 
Here, we demonstrate this capability by realizing a binary mixture of strongly-interacting excitons in high-lying quantum states with different principal quantum numbers via two-color optical excitation in Cu$_2$O. Through an adequate choice of Rydberg states and laser intensities, we implement an asymmetric Rydberg blockade \cite{saffman2008} between excitons, where the interactions among identical excitons are of minor importance while the interactions between excitons in different quantum states result in strong spatial correlations and an extended excitation blockade of inter-species exciton pairs. Our experiments exploit this asymmetry to employ one exciton species as a probe for the presence of the other, and reveal direct spectroscopic signatures of the emerging spatial correlations between the two species as well as the distinct power-law character of their mutual interaction. 

\begin{figure*}[t!]
\begin{center}
\includegraphics[width=0.67\textwidth]{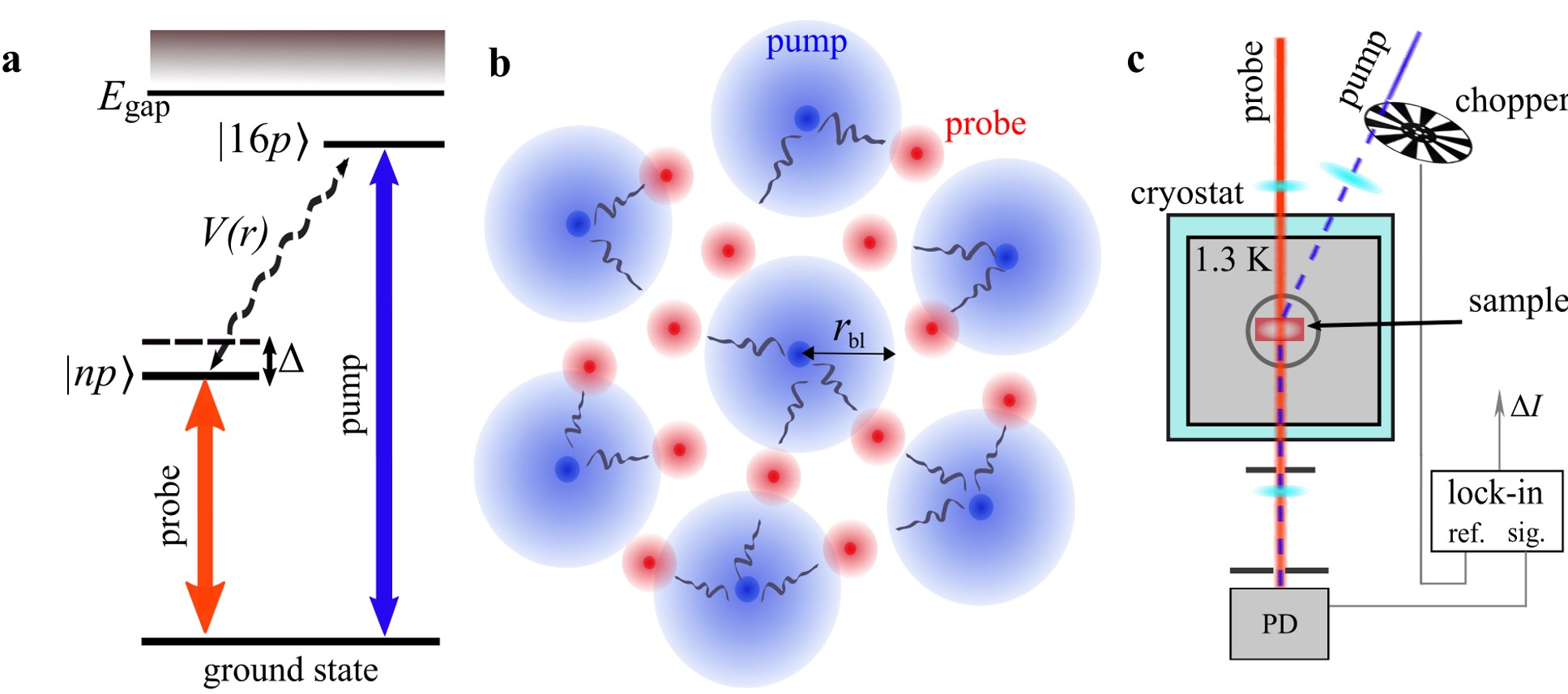} 
\includegraphics[width=0.67\textwidth]{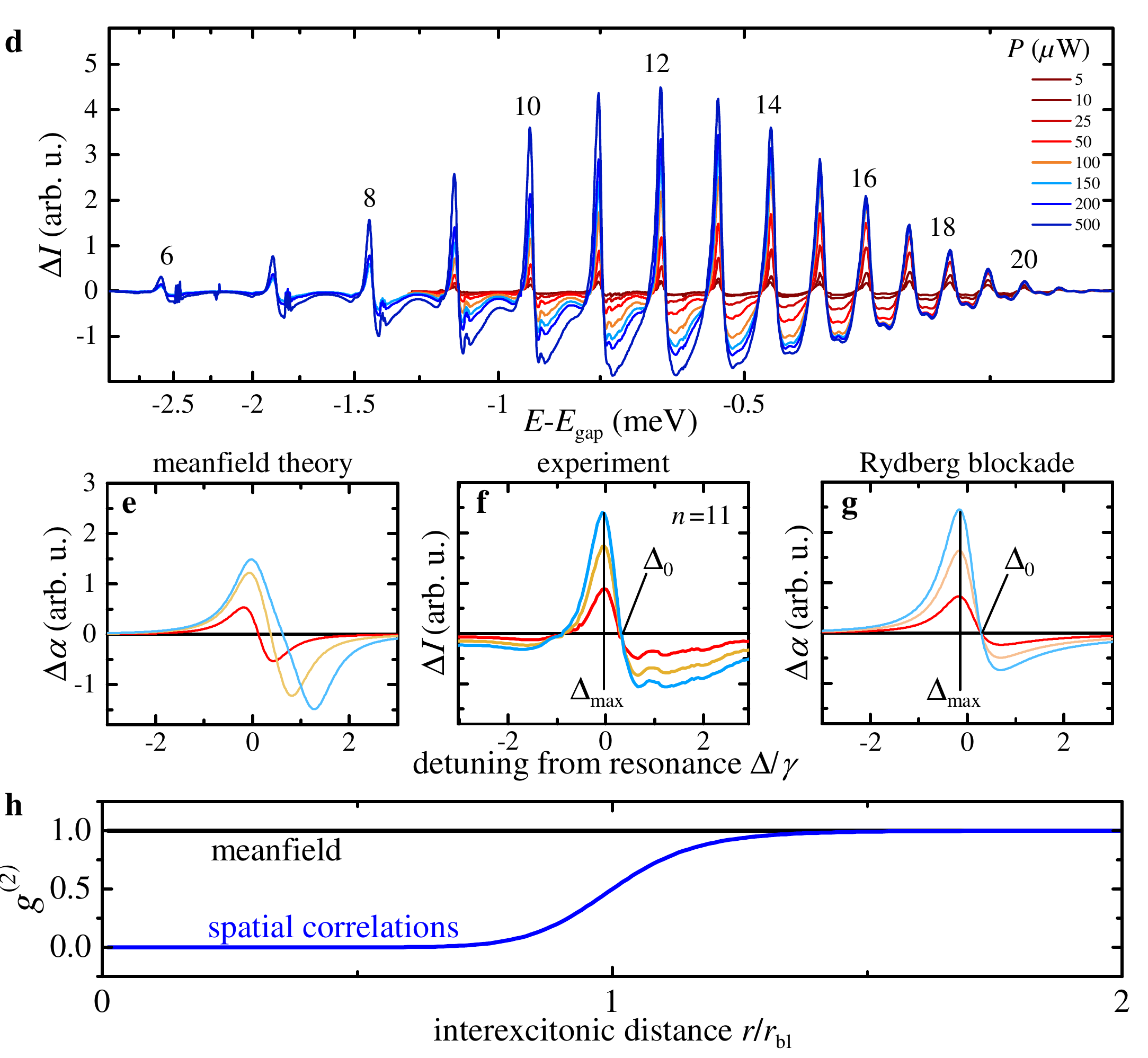} 
\end{center}
\caption{{Two-color spectroscopy of excitonic Rydberg series.}
\textbf{a},~A weak pump laser (blue) excites excitons in the $16p$ state which interact with $np$ excitons that are excited by a weak probe laser (red). 
\textbf{b},~These inter-state interactions have a profound effect on the probe-exciton dynamics and tend to block their optical generation within pump-probe distances below the blockade radius $r_{\rm bl}$.
\textbf{c},~In our experiment, we modulate the pump beam by an optical chopper and use a lock-in amplifier to lock the recording of the transmitted probe photons to the modulation frequency. In this way, the output signal of the photodiode (PD) that collects the probe photons is directly proportional to the pump-induced differential probe transmission $\Delta I$. 
Examples of the recorded spectra are shown in panel \textbf{d}, which display a series of clear resonances for probe excitons with principal quantum numbers from $n=6$ to $n=20$. The increasing signal strength with the pump power, $P$,  provides indication for the presence of interactions between pump and probe excitons.
A common meanfield treatment of such interactions (panel \textbf{e}), however, fails to reproduce the qualitative features of the measured resonances, as shown in panel \textbf{f}. On the other hand, the observed universal maximum and root at fixed laser detunings $\Delta_{\rm max}$ and $\Delta_0$, respectively, are well explained by a theory that accounts for strong exciton correlations and excitation-blockade effects (panel \textbf{g}). The underlying spatial correlation function, $g^{(2)}(r)$ between pump and probe excitons, shown in panel \textbf{h}, displays an extended exciton blockade for distances below the blockade radius $r_{\rm bl}$, which is on the order of several $\mu$m in our experiments.  }
\label{fig1}
\end{figure*}

The asymptotic interaction between neutral particles, such as atoms or excitons, is dominated by the van der Waals potential that decreases as a simple power law, $V(r)=C_6/r^6$, with the inter-particle distance $r$. For electronic ground states, however, the van der Waals coefficient $C_6$ is typically very small and short-distance exchange effects play an important role in the overall interaction that can often be described in terms of zero-range collisions \cite{Ciuti1998}. The drastic increase of $C_6\sim n^{11}$ with the principal quantum number $n$ of excited states, on the other hand, gives rise to exaggerated van der Waals interactions that can be sufficiently strong to even affect the very process of optically generating high-lying Rydberg states, as observed and exploited in cold-atom systems. Hereby, the presence of one excited particle can perturb or even prevent the excitation of another by shifting its energy via the interaction between the two Rydberg states. This Rydberg blockade \cite{lukin} not only enables rapid saturation at very low light intensities \cite{pritchard_cooperative_2010,sevincli_nonlocal_2011,peyronel_quantum_2012}, but can also lead to the emergence of strongly correlated many-body states of Rydberg excitations \cite{Pohl2010,Schauss2015}. While such correlations can be observed directly on a microscopic level in cold-atom experiments \cite{Schauss2012,barredo_demonstration_2014}, they are more difficult to access in solid-state systems. 

Our experiments make it possible to probe blockade effects in a semiconductor material at an extreme range of several $\mu$m. To this end, we have implemented a two-color excitation scheme, whereby two spectrally-narrow laser beams with different frequencies excite Rydberg excitons with different principal quantum numbers $n$ and $n'$, respectively (see Fig.~\ref{fig1}\textbf{a}). We use a natural Cu$_2$O crystal that is cut and polished to a thickness of 30 $\mu$m and held at a temperature of 1.3~K. 
The selection rules associated with the involved band symmetries allow us to excite excitonic $p$-states in the so-called yellow series via single-photon absorption in the optical domain. 
Hereby, a pump laser generates $p$-state excitons with $n'=16$, while another weak probe field with variable frequency creates Rydberg excitons with variable principle quantum numbers $n=6,...,20$ that sense the presence of the pump excitons via their mutual interactions (Fig.~\ref{fig1}\textbf{b}). By varying the power of the pump beam, $P$, we can control the density of the pump excitons and monitor their effect on the probe-field absorption around a given probe-exciton resonance. 

In order to achieve the required sensitivity for an accurate measurement of interaction effects, we modulate the pump laser by an optical chopper and detect the transmitted intensity of the probe laser, $I$, with a photodiode that is connected to a lock-in amplifier and locked to the pump modulation frequency (see Fig.~\ref{fig1}\textbf{c} and the Methods for details). By choosing a low modulation frequency of 3.33~kHz far below any relevant dynamical frequency scale in the system, we ensure that the resulting signal yields the spectral form of the pump-induced change in the probe transmission $\Delta I\propto I(P)-I(P=0)$.  Importantly, this approach makes it possible to scan the Rydberg series of the probe excitons while maintaining otherwise stable excitation conditions.

Figure~\ref{fig1}\textbf{d} summarizes the result of such measurements and shows differential probe spectra for a varying power $P$ of the pump laser (see Supplementary Information~I for further details). Since the density, $\rho$, of pump excitons grows with $P$, the depicted power dependence carries information about the effects of interactions between pump and probe excitons. A positive signal corresponds to an increased transmission caused by the generated pump excitons, and the observed transmission peaks can be assigned to the Rydberg-state resonances of the probe excitons, as indicated in Fig.~\ref{fig1}\textbf{d}. For each individual peak, we observe a linear growth of the signal with $P$ but no measurable shift of the position, $\Delta_{\rm max}$, of the transmission peaks, see closeup in Fig~\ref{fig1}\textbf{f}. At first glance, this comes as a surprise, as an increasing density of pump excitons would be expected to cause a larger shift of the probe-exciton line, similar to the known behavior of quantum-well excitons with short-range interactions \cite{butov_magneto-optics_1999, laikhtman_exciton_2009}. In addition, we find that the differential transmission $\Delta I$ crosses zero at the high-energy side of the resonance at a detuning $\Delta_0$ that is entirely independent of the pump-laser power. Indeed all of these features provide strong indication for the emergence of extended spatial correlations during the optical generation of excitons, as we shall see below. 

Let us first assume that spatial correlations are insignificant. In this case, the interactions between pump and probe excitons would lead to a simple energy shift, $\Delta_{\rm mf}=\rho\int {\rm d} {\bf r} V({ r})$, of the probe exciton resonance that is determined by the exciton interaction potential $V({ r})$ and increases linearly with the pump-exciton density $\rho$. 
Note that such a meanfield treatment describes many aspects of excitons with weak short-range interactions \cite{carusotto_quantum_2013}, such as the observed nonlinear spectral properties of semiconductor microcavities \cite{kasprzak_boseeinstein_2006}, or the fluid-like behavior of exciton-polaritons in such settings \cite{amo_superfluidity_2009}. For our experiments, however, a pure meanfield picture fails to capture the essential physics of the exciton dynamics as it predicts a power-dependent position of the maxima and roots of the differential transmission. The meanfield prediction is illustrated in Fig.~\ref{fig1}\textbf{e} for three different interaction strengths, and stands in stark contrast to our measurements shown in the middle panel of the same figure. 

Such qualitative discrepancies indicate that emerging correlations between the interacting excitons play a significant role during their optical generation. We can explore this further by theoretically considering the correlated excitation dynamics of probe excitons in a background of pump excitons with a density $\rho$. As the interaction, $V(r)$, between the pump- and probe-excitons shifts the energy of exciton-pair states, it leads to strong spatial correlations following a pump-probe correlation function, that assumes a particularly simple form 
\begin{equation}\label{eq:g2}
g^{(2)}(r)=\frac{\gamma^2/4+\Delta^2}{\gamma^2/4+[V(r)-\Delta]^2}
\end{equation}
to lowest order in the probe intensity (see Supplementary Information~III and IV). The function $g^{(2)}(r)$ yields the probability to generate a probe exciton in the vicinity of a pump exciton at a distance $r$. As illustrated in Fig.~\ref{fig1}\textbf{h}, this probability vanishes rapidly as the distance between excitons falls below a critical radius $r_{\rm bl}$ that is determined by the linewidth $\gamma$ and frequency detuning $\Delta$ (see Fig.~\ref{fig1}\textbf{a}) of a given Rydberg-state resonance. In our experiments, the range of this exciton blockade can take on remarkably large values of several $\mu$m that would significantly affect the probe-beam transmission as expressed by the change of the absorption coefficient
\begin{equation}\label{eq:abs}
\alpha=\alpha_0\left(1-\rho\int\!\frac{(V(r)-2\Delta)V(r)}{\gamma^2/4+\Delta^2}g^{(2)}(r){\rm d}{\bf r}\right)
\end{equation}
relative to the absorption $\alpha_0$ in the absence of the pump-beam excitons. For low absorption, the proportionality $\Delta I\propto \Delta \alpha=\alpha_0-\alpha$ affords direct comparisons of our differential transmission measurements with the prediction of equations~\eqref{eq:g2} and \eqref{eq:abs}. As demonstrated in Fig.~\ref{fig1}\textbf{g}, our theory for the correlated exciton dynamics indeed reproduces the essential features of our observations. In particular, it yields a maximum and a common root on the blue side of the differential transmission spectrum that is independent of the pump-beam intensity. This characteristic behavior is also evident from equation~\eqref{eq:abs}, since the spectral shape of $\alpha_0(\Delta)-\alpha(\Delta)$ is solely determined by the exciton interaction and pair correlation function, while the pump-exciton density merely enters as a linear pre-factor, which reproduces the observed overall linear scaling of the signal with the pump-beam power.

While these characteristic features already provide clear evidence for the Rydberg blockade and emergence of strong exciton correlations, we can also obtain more detailed information about the underlying van der Waals interaction by analyzing the maxima of the differential transmission around each Rydberg-state resonance. To this end, we have recorded the pump-power dependence of the maximum signal strength, which exhibits a linear scaling $\Delta I=\beta P$ for low pump-beam intensities with a slope $\beta(n,n^\prime)$ that depends on the principal quantum number of both involved Rydberg states (see Supplementary Information~II and V). By evaluating the integral in equation~\eqref{eq:abs} for near-resonant excitation, $\Delta/\gamma\ll1$, one finds that the slope should scale as $\beta\propto r_{\rm bl}^3$. Here, the blockade radius $r_{\rm bl}=\left(\frac{C_6}{\gamma/2}\right)^{1/6}$ corresponds to the distance below which the van der Waals interaction $V(r_{\rm bl})=C_6/r_{\rm bl}^6$ starts to exceed the width of the probe-exciton resonance. By scaling the observed slope with the measured exciton linewidths, we can thus probe the state-dependence of the van der Waals coefficient, $C_6\propto \beta^2\gamma$. 
\begin{figure}[t!]
	\begin{center}
		\includegraphics[width=1\columnwidth]{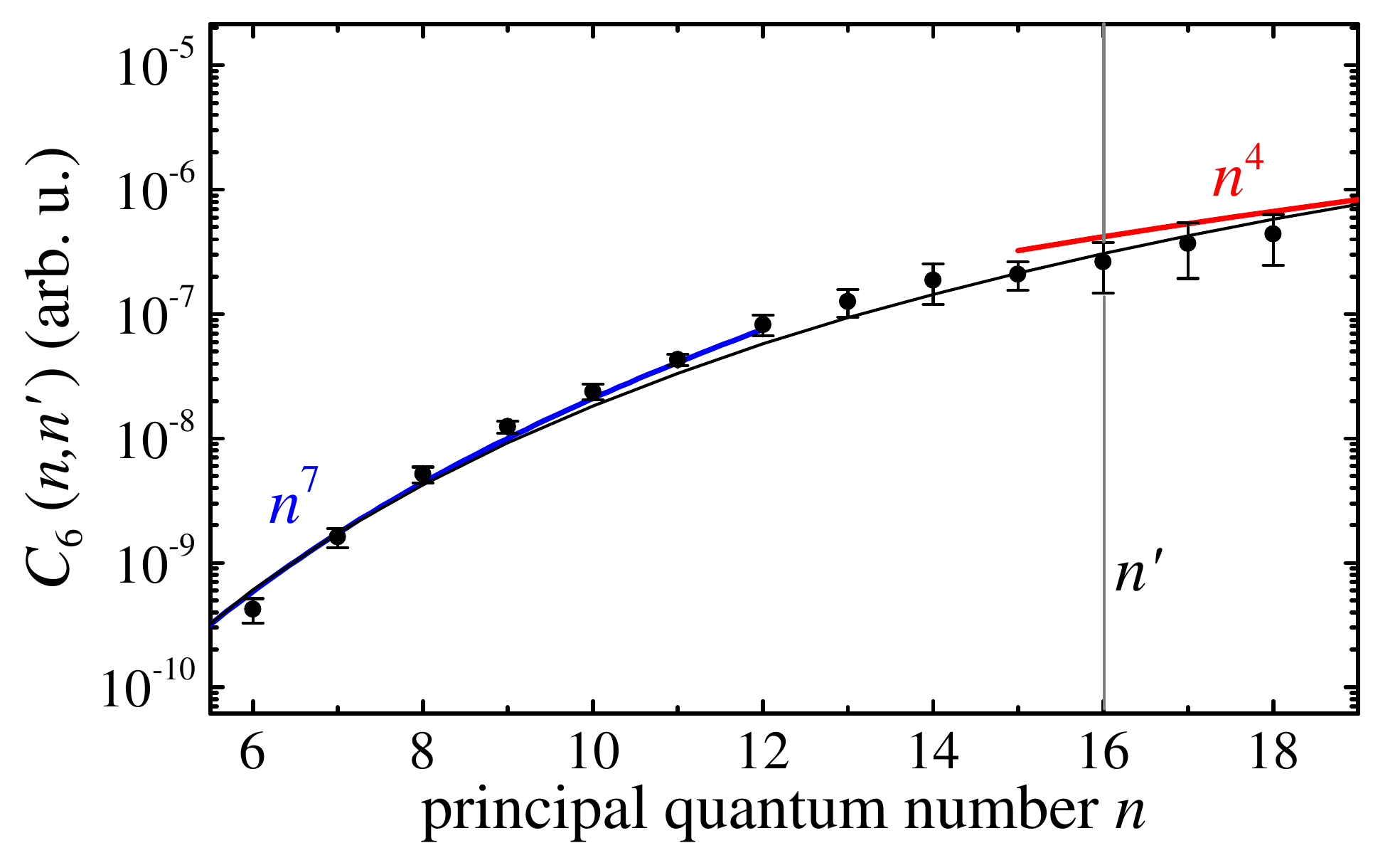} 
	\end{center}
	\caption{{Observed Rydberg scaling of the van der Waals coefficient.} 
		Measured values (black dots) and theoretical predictions (black line) for the inter-state van der Waals coefficient $C_6(n,n^\prime)$, that yields the strength of long-range interactions between excitons with different principal quantum numbers $n$ and $n^\prime$. Here, $n^\prime=16$ is held fixed while varying $n$, whereby the theory curve has been scaled to match the experiment at $n=18$. For $n\ll n^\prime$, theory and experiment give a  simple $\sim n^7$ scaling (blue), which flattens in the opposite limit, when approaching the expected $\sim n^4$ behavior for $n\gg n^\prime$. 
		}
	\label{fig2}
\end{figure}
Our experimentally determined values are shown in Fig.~\ref{fig2} and indicate a rapid increase of the interaction strength by about three orders of magnitude over the probed range of principal quantum numbers. From the characteristic scaling laws for the level spacings and transition matrix elements of hydrogenic Rydberg states (see Supplementary Information~V) one would expect a characteristic scaling of the interstate van der Waals interaction as 
\begin{equation}
C_6=n^7n'^{ 4},
\end{equation}
if $n\ll n'$. Indeed, our measurements show such a rapid $n^7$-increase of the interaction for low-lying states ($n<n^\prime$) and also indicate a cross-over to a slower increase, $\sim n^4$, once the excitation level of the probe-exciton exceeds that of the pump excitons at $n^\prime=16$ in accordance with the model prediction. 

\begin{figure}[t!]
	\begin{center}
		\includegraphics[width=\columnwidth]{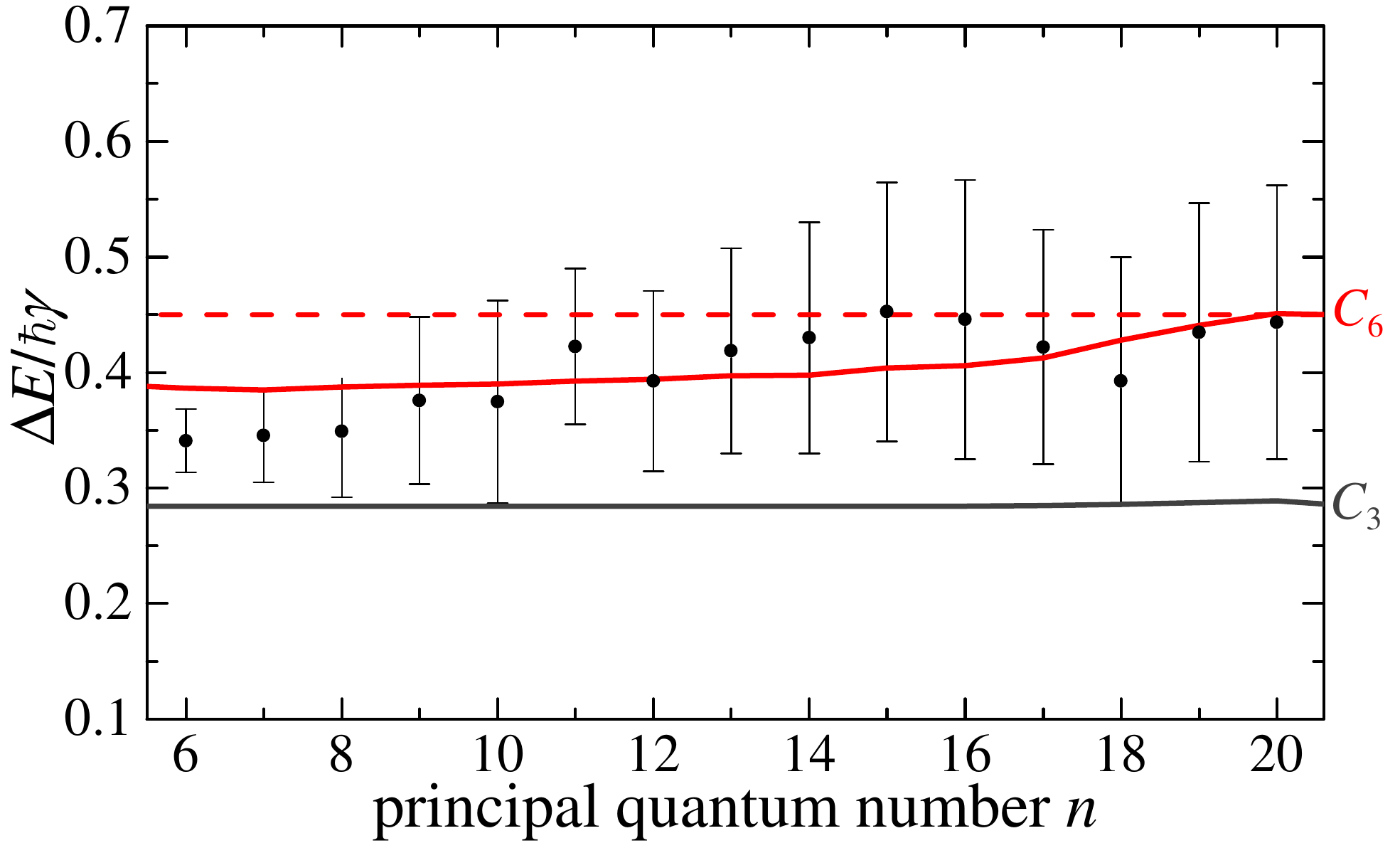}
	\end{center}
	\caption{
	{Universal behavior.} The scaled difference $\Delta E/\hbar\gamma=(\Delta_0-\Delta_{\rm max})/\gamma$ between the maxima and roots of the transmission signal does not depend on the pump intensity (see Fig.1\textbf{f} and \textbf{g}) and is shown here as a function of the principal quantum number $n$ of the probe excitons. For homogeneously broadened laser excitation, it becomes a universal quantity that only depends on the shape of the interaction and is given by $\Delta E/\hbar\gamma=0.45$ for van der Waals interactions, $V(r)=C_6/r^6$, as indicated by the dashed red line. Phonon coupling leads to asymmetric line broadening and causes slight deviations from this behavior (red solid line) in excellent agreement with the measurements shown by the black dots. Other potentials, such as direct dipole-dipole interactions, $V(r)=C_3/r^3$, do not match the observations (grey line). }
	\label{fig3}
\end{figure}

The peak height of the differential transmission, therefore, demonstrates the power-law scaling of the interaction strength with the principal quantum number of the excitons. In addition, however, its spectral shape also carries spatial  information about the power-law decay of the interaction potential as a function of the distance $r$ between the excitons. 
In fact, equations~\eqref{eq:g2} and \eqref{eq:abs} predict that the scaled energy difference $\Delta E/\hbar\gamma=(\Delta_0-\Delta_{\rm max})/\gamma$ between the positions of the maximum and zero crossing of the transmission signal (see Fig.~\ref{fig1}\textbf{c}) should depend only on the form and strength of the interaction potential. For pure power-law potentials, $\Delta E/ \hbar\gamma$ even becomes independent of the interaction strength and assumes a characteristic universal value for each power-law exponent, which is given by $\Delta E/ \hbar\gamma=0.45$ for the van der Waals interaction, $V(r)\sim1/r^6$. As shown in Fig.~\ref{fig3}, our measurements indeed approach this universal value for high principal quantum numbers $n$ of the Rydberg excitons. The spectral line shape of lower lying states is stronger affected by exciton-phonon coupling, which leads to asymmetric broadening that causes small deviations from the universal behavior. These deviations are remarkably well captured by corrections based on Fano-resonance theory \cite{toyozawa_interband_1964} as shown by the red line in Fig.~\ref{fig3} (see Supplementary Information~VI). 

The combined analysis of our pump-probe measurements thus offer broad insight into the microscopic mechanisms of exciton interactions, and provide a first experimental case for the action of long-range electrostatic van der Waals interactions between excitons in a semiconductor. Being sensitive to the shape and strength of the underlying interaction potential, our scheme makes it possible to discriminate between different types of interactions and clearly excludes direct dipole-dipole interactions with $V(r)\sim1/r^3$ (see Fig.~\ref{fig3}). One can explore this capability further by pumping the system above the bandgap to create free charges instead of initial excitons (see Supplementary Information~VII). 
The resulting pump-probe spectra give clear evidence for a rapid dynamical recombination of the produced electron-hole plasma into Rydberg-exciton states similar to the behavior of ultracold atomic plasmas \cite{killian_formation_2001, Takahata2018}. 
Such elementary relaxation processes could become directly accessible to real-time measurements by operating the presented pump-probe approach with short laser pulses. Indeed, this offers an exciting outlook on the developed method, which, when operated below the bandgap, would open a new experimental window into the non-equilibrium dynamics of strongly interacting excitons in a semiconductor. Such temporal control could also make advanced studies of quantum mixtures possible, and, for example, enable the investigation of impurity physics and formation of exotic polarons \cite{camargo2018} in mixtures of ground- and Rydberg-state excitons. Generally, the demonstrated ability to enhance interactions and realize spatially extended blockade effects between different exciton states, while maintaining weak intra-state nonlinearities, suggests interesting applications, such as highly efficient few-photon switches \cite{chang_quantum_2014} that could be implemented with optical resonators in near-term experiments. 

\clearpage

\section*{Methods}
\subsection*{Experimental Setup}
We use a two-color pump-probe setup employing two stabilized dye-lasers with a narrow linewidth of 5~neV that serve as pump and probe beams. 
The studied sample is a Cu$_2$O slab with a thickness of $L=34~\mu$m cooled down to 1.35\,K inside a liquid helium bath and mounted free of strain. 
Both lasers are in perfect spatial overlap. To assure a homogeneously distributed pump exciton density, the pump laser's beam waist is set to 300~$\mu$m while the probe's waist is set to 100~$\mu$m, both measured at full width at half maximum. 
The probe power is kept as low as 1 $\mu$W to ensure negligible interactions among probe excitons. 
The pump beam is periodically switched on and off by an optical chopper blade with a frequency of 3.33~kHz. 
The transmitted probe intensity is detected with a photodiode connected to the signal input of a lock-in amplifier (see Fig.~\ref{fig1}\textbf{c}). The reference signal is provided by the optical chopper and contains the modulation frequency of the pump beam. The lock-in amplifier mixes both signals which yields a frequency-independent part that is proportional to the pump-induced change in the probe beam transmission $\Delta I$. Due to the used low modulation frequency this serves as a quasi-CW signal. 

\bibliographystyle{naturemag} 
\bibliography{RydbergBib}

\clearpage
\noindent{\bf Acknowledgements} We would like to thank Sjard Ole Krüger for fruitful discussions. \\
This work was supported by the Carlsberg Foundation through the 'Semper Ardens' Research Project QCooL, by the DFG through the SPP 1929 GiRyd (project numbers: 316133134, 316159498 and 316214921), by the European Commission through the H2020-FETOPEN project ErBeStA (No. 800942), by the DNRF through a Niels Bohr Professorship to TP and the DNRF Center of Excellence “CCQ” (Grant agreement no.: DNRF156) and by the NSF through a grant for the Institute for Theoretical Atomic, Molecular, and Optical Physics at Harvard University and the Smithsonian Astrophysical Observatory.

\end{document}


\title{\huge \bf{Asymmetric Rydberg blockade of giant excitons in Cuprous Oxide\\
	Supplementary Information}}
	
	\author{Julian Heck\"otter}
	\thanks{These two authors contributed equally }
	\affiliation{%
		Experimentelle Physik 2,
		Technische Universit\"at Dortmund,
		D-44221 Dortmund, Germany
	}%
	\author{Valentin Walther}%
	\thanks{These two authors contributed equally }
	\affiliation{Center for Complex Quantum Systems, Department of Physics and Astronomy, Aarhus University, Ny Munkegade 120, DK-8000 Aarhus C, Denmark
	}%
	\affiliation{ITAMP, Harvard-Smithsonian Center for Astrophysics, Cambridge, Massachusetts 02138, USA}
	\author{Stefan Scheel}
	\affiliation{
		Institut f\"ur Physik, Universit\"at Rostock, Albert-Einstein-Stra{\ss}e 23, D-18059 Rostock, Germany
	}
	\author{Manfred Bayer}
	\affiliation{%
		Experimentelle Physik 2,
		Technische Universit\"at Dortmund,
		D-44221 Dortmund, Germany
	}%
	\affiliation{Ioffe Institute,
	Russian Academy of Sciences,
	194021 St. Petersburg, Russia}

	\author{Thomas Pohl}%
	\affiliation{Center for Complex Quantum Systems, Department of Physics and Astronomy, Aarhus University, Ny Munkegade 120, DK-8000 Aarhus C, Denmark
	}%
	\author{Marc A{\ss}mann}
	\affiliation{%
		Experimentelle Physik 2,
		Technische Universit\"at Dortmund,
		D-44221 Dortmund, Germany
	}%
	
	\email{marc.assmann@tu-dortmund.de}
	
	\date{\today}
	
	\maketitle

\tableofcontents

\section{Experimental details: Separation of interaction regimes} \label{sec:correction}

\begin{figure}[htbp] 
	\centering
	\includegraphics[width=0.85\columnwidth]{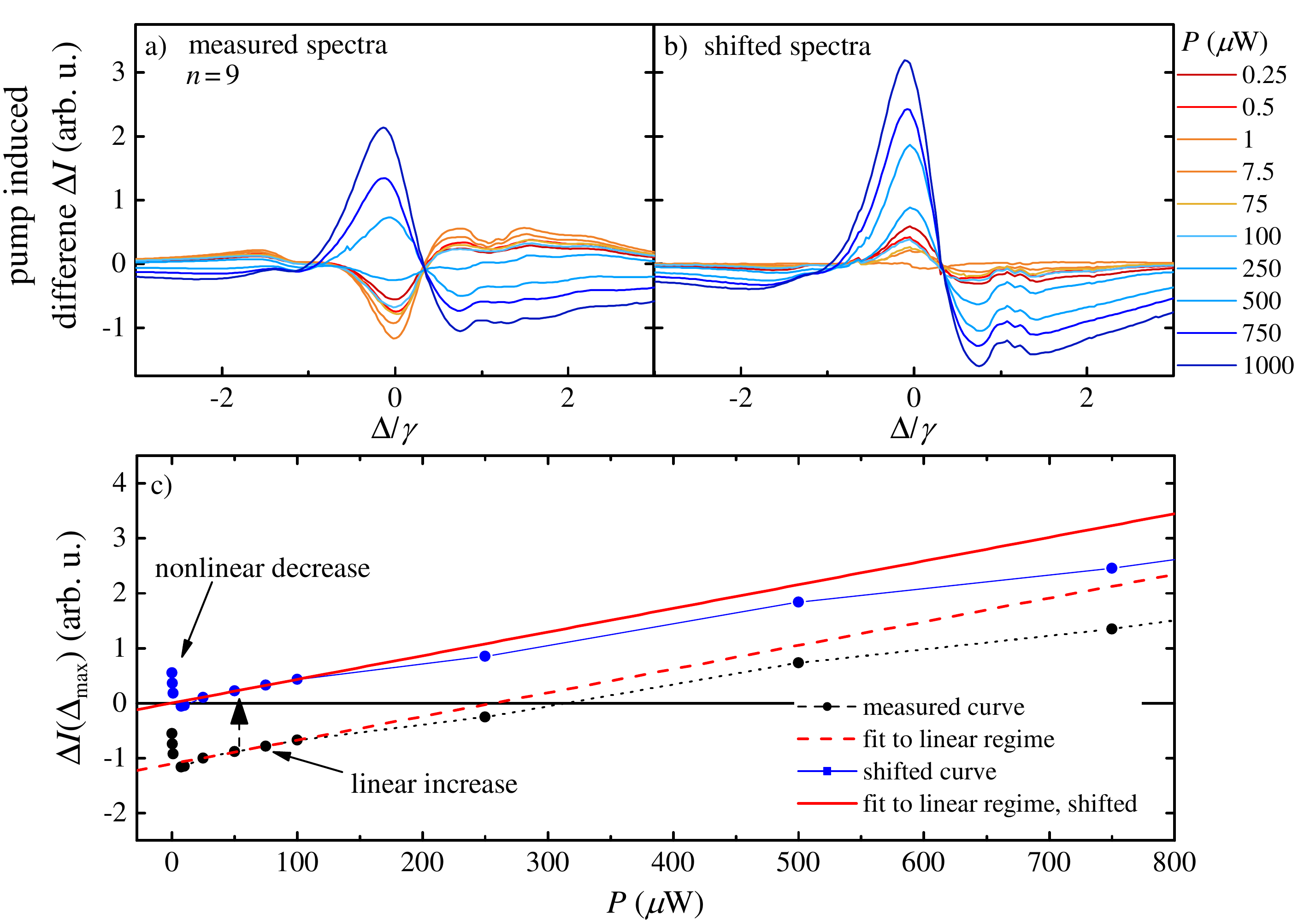}
	\caption{{Data shift:} 
		a) Uncorrected differential transmission spectra around the $n=9$ resonance. Here, we find negative peak amplitudes for low excitation powers that correspond to a decrease of transmission. 
		b) Corrected spectra around $n=9$ peak after subtraction of the initial nonlinear decrease of transmission. 
		c) 
		Black dots show the measured maximum signal amplitude of the resonance in panel~a) as a function of pump power. The Rydberg interaction is extrapolated linearly towards low intensities (red dashed line). 
		The obtained offset is subtracted from the data resulting in the blue curve. The linear slope does not change (red solid line). 
	}
	\label{fig:SI1}
\end{figure}
The data presented in Fig.~1 in the main text and in Supplementary Fig.~\ref{fig:SI4} originally show a defect-induced initial nonlinear decrease of transmission for low pump powers that saturates fast. 
In order to focus on the Rydberg interaction regime, the data is corrected for this nonlinearity, as described in the following exemplarily for the $n=9$ resonance. 
Supplementary Fig.~\ref{fig:SI1}a) shows the measured differential transmission $\Delta I$ around the $n=9$ resonance for different pump powers $P$ with a focus on the low power range. 
The pump laser energy is set to the $n'=16$ resonance. 

Starting at low powers (dark red), the transmission around the resonance first decreases, resulting in increasing negative values of the differential transmission $\Delta I$. With growing pump power, this trend is slowed (orange) and finally reversed into enhanced transmission ($\Delta I>0$, light blue). 
For a more quantitative discussion, the maximum signal amplitude of the resonance in panel~a), $\Delta I(\Delta_\text{max})$, is shown as a function of pump power by the black dotted line in panel~c). 
As indicated by the arrow, the nonlinear decrease of transmission can be clearly seen for low pump powers and is found to saturate fast at powers around 5~$\mu$W in this particular case. 
After saturation, the nonlinearity is directly followed by a linear increase of transmission over a wider range of powers. 
This linear increase is indicated by the red dashed line (see below) and describes the expected linear optical response of the probed state due to Rydberg interactions for low pump beam intensities. 
At higher powers, the curve flattens and finally saturates. 

In order to separate the initial nonlinear decrease of transmission from the exciton interaction regime, the data is extrapolated to zero pump power in the range of linear power dependence, shown by the red dashed line, and the resulting intercept is subtracted from the spectra. The corrected data is given by the blue line with an unchanged slope, given by the red solid line. 
This procedure is repeated for the whole dataset at every detuning. 
The resulting corrected spectra around the resonance $n=9$ are shown in Supplementary Fig.~\ref{fig:SI1}b).

\section{Experimental details: Extraction of $\beta$}
The maximum difference signal at each resonance, $\Delta I (\Delta_\text{max})$, exhibits a linear dependence on pump power $P$ in the range of low powers with a characteristic slope $\beta$ that depends on the principal quantum number $n$ of the probed state. 
To evaluate the $n$-dependent scaling of $\beta$, the maximum signal amplitude of each resonance is extracted from the data in Fig.~1 in the main text as a function of pump power. 
The obtained values are shown in Supplementary Fig.~\ref{fig:SI3} for the states with quantum numbers $n=6$ to $n=18$. 
The linear slope $\beta$ is obtained by fits to the data within the range of a linear power dependence, as shown by the solid lines. 
The slope grows with increasing principal quantum number $n$ of the probed state, reflecting a larger interaction strength $C_6$ for Rydberg excitons with higher $n$, see Eqs.~\eqref{eq:beta} and \eqref{eq:C6} in Supplementary Note \ref{sec:beta}.  
Consequently, the upper end of the linear power range and the onset of saturation move to smaller pump powers with increasing $n$. 

\begin{figure}[b]
	\centering
	\includegraphics[width=0.72\textwidth]{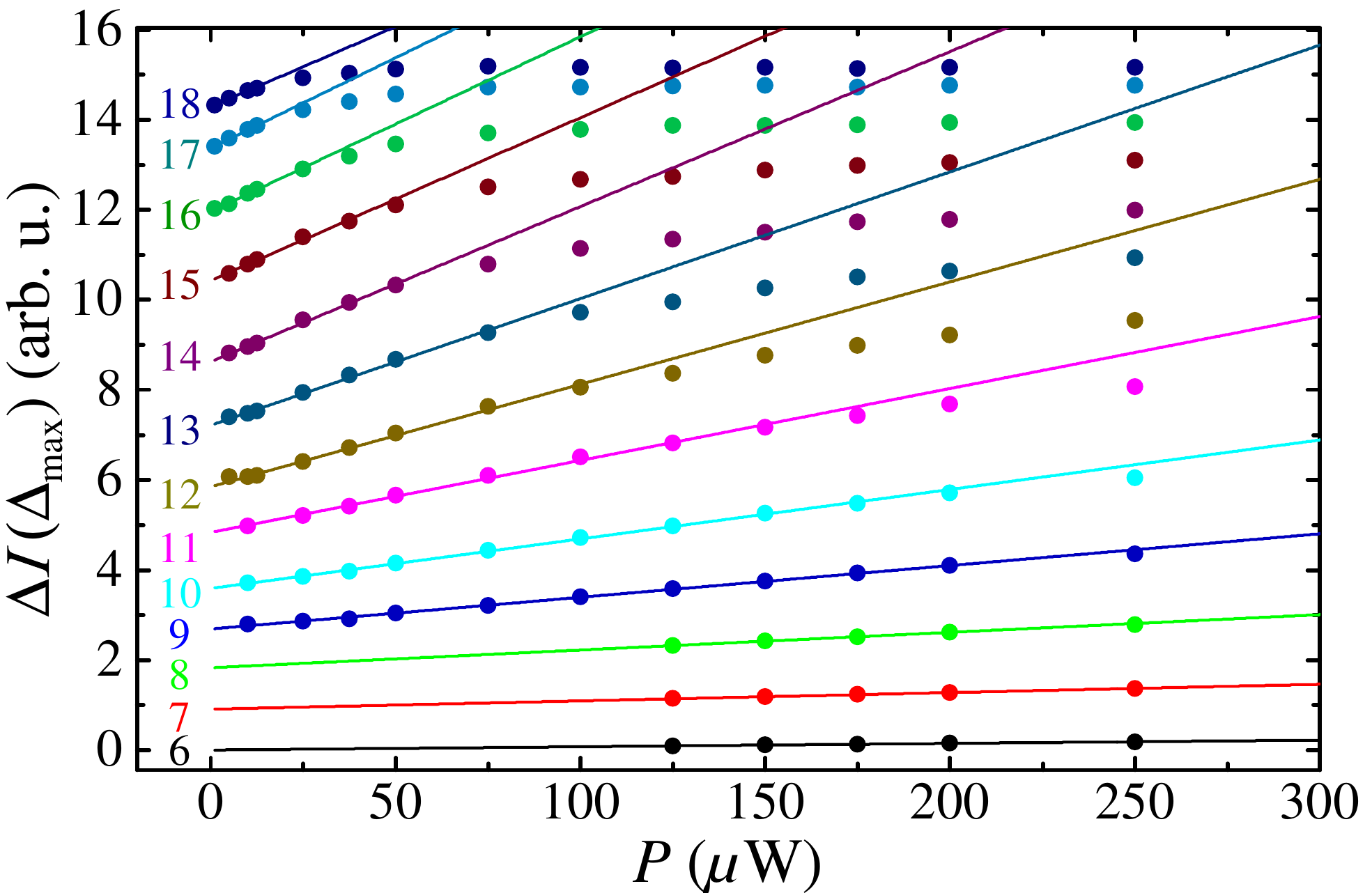}
	\caption{{Characteristic scaling (${n'=16}$):} Maximum differential signal $\Delta I(\Delta_\text{max})$ as a function of pump power $P$ for principal quantum numbers $n=6$ to $n=18$, obtained from the data shown in Fig.~1d in the main text. 
	The solid lines show fits with a slope $\beta$ in the range of a linear dependence on pump power for each resonance. 
	The slope of each fit increases with $n$ reflecting a stronger optical response for states with larger $n$. }
	\label{fig:SI3}
\end{figure}

\clearpage

\section{Exciton correlations} \label{sec:ex_corr}
Here, we present details of the theory of the coupled dynamics of pump and probe excitons. Probe excitons, described by $\hat{X}(\vec{r})$, are created at a rate $g$ from the coherent probe laser of amplitude $\mathcal{E}$ with a detuning given by $\Delta=\omega_\text{in}-\omega_\text{ex}$. The excitonic resonance has a linewidth $\gamma$. 
Interactions with the pump excitons, described by $\hat{Y}(\vec{r})$, are captured by a potential $V$. We assume that the density of probe excitons is much lower than the pump exciton density, such that interactions among the probe excitons can be neglected.
The Heisenberg equation of motion for a probe exciton describes the dynamics
\begin{equation}\label{eq:motion_mod}
\begin{aligned}
\partial_{t} \hat{X}(\vec{r}) &= -\frac{\Gamma}{2}\hat{X}(\vec{r}) - i g\mathcal{E}(\vec{r}) - i \int d \vec{r}^{\prime} V(|\vec{r}-\vec{r}^{\prime}|)\hat{Y}^{\dagger}(\vec{r}^{\prime})\hat{Y}(\vec{r}^{\prime})\hat{X}(\vec{r}),
\end{aligned}
\end{equation}
where decay and detuning are summarized into the complex linewidth $\Gamma=\gamma-2i\Delta$.
To obtain a solution for the polarization, $\propto \langle \hat{X} \rangle$, we formulate a hierarchy of exciton correlators based on Eq.~(\ref{eq:motion_mod}) and subsequently solve the steady-state equations
\begin{equation} \label{eq:coherence}
 \begin{aligned}
 \partial_t \hat{Y}^\dagger(\vec{r}')\hat{Y}(\vec{r}')\hat{X}(\vec{r})
=&-\frac{\Gamma}{2}\hat{Y}^\dagger(\vec{r}')\hat{Y}(\vec{r}')\hat{X}(\vec{r})-ig\hat{Y}^\dagger(\vec{r}')\hat{Y}(\vec{r}')\mathcal{E}(\vec{r}) \\
&-iV(|\vec{r}-\vec{r}'|)\hat{Y}^\dagger(\vec{r}')\hat{Y}(\vec{r}')\hat{X}(\vec{r}) \\
&-i\int d\vec{r}'' V(|\vec{r}-\vec{r}''|)\hat{Y}^\dagger(\vec{r}')\hat{Y}^\dagger(\vec{r}'')\hat{X}_p(\vec{r}')\hat{X}(\vec{r}'')\hat{X}(\vec{r}).
 \end{aligned}
\end{equation}
In the derivation of Eq.~(\ref{eq:coherence}), we neglect the time dependence of the pump exciton density, as the pump laser is chopped and the pump exciton lifetime is long. 
The second-to-last term describes the interaction-induced shift inflicted on the probe exciton coherence at $\vec{r}$ by the presence of a pump exciton at position $\vec{r}^\prime$, while the last term describes simultaneous interactions of a probe exciton with multiple pump excitons. In the limit of low densities relevant for the experiments, we can drop the last term, thus closing the system of equations after taking expectation values 
\begin{align}
 \langle \hat{Y}^\dagger(\vec{r}')\hat{Y}(\vec{r}')\hat{X}(\vec{r}) \rangle = -i   \frac{g \rho(\vec{r}')}{\Gamma/2 + iV(|\vec{r}-\vec{r}^\prime|)} \mathcal{E}(\vec{r}), 
\end{align}
where $\rho(\vec{r}') = \langle \hat{Y}^\dagger(\vec{r}')\hat{Y}(\vec{r}')\rangle$ is the pump exciton density. 
This leads to the steady-state expression for the exciton coherence
\begin{align}\label{eq:exp_X}
 \langle \hat{X}(\vec{r}) \rangle = -\frac{2ig}{\Gamma} \mathcal{E}(\vec{r}) \left( 1 - i \int d \vec{r}^{\prime}  \rho(\vec{r}') \frac{V(|\vec{r}-\vec{r}^{\prime}|) }{\Gamma/2 + iV(|\vec{r}-\vec{r}^\prime|)} \right).
\end{align}
The expectation value for finding a probe exciton in the vicinity of a pump exciton can be calculated analogously from Eq.~(\ref{eq:motion_mod})
\begin{equation}
 \begin{aligned}
 \langle \hat{X}^\dagger(\vec{r}) \hat{Y}^\dagger(\vec{r}')\hat{Y}(\vec{r}')\hat{X}(\vec{r}) \rangle &= \frac{2}{\gamma} \Re\left[ig \langle\hat{Y}^\dagger(\vec{r}')\hat{Y}(\vec{r}')\hat{X} (\vec{r}) \rangle \mathcal{E}^*(\vec{r})\right] \\
 &= \frac{g^2 \rho^2(\vec{r})}{\gamma^2/4+\Delta^2} g^{(2)}(\vec{r},\vec{r}^\prime) |\mathcal{E}(\vec{r})|^2,
\end{aligned}
\end{equation}
where we define the probe-pump correlation function
\begin{align}
 g^{(2)}(\vec{r},\vec{r}^\prime) = \frac{\gamma^2/4+\Delta^2}{\gamma^2/4+(V(|\vec{r}-\vec{r}^\prime|)-\Delta)^2},
\end{align}
as given in Eq.~(1) of the main text.

The above theory is qualitatively different from a meanfield model that neglects correlations between the pump and probe excitons. Thus, the latter finds a flat correlation function $g^{(2)}$, as shown in Fig.1h of the main text. In the meanfield, the coherence is directly obtained by factorizing the last term in Eq.~(\ref{eq:motion_mod})
\begin{align} \label{eq:coherence_mf}
 \langle \hat{X}(\vec{r}) \rangle = - i \frac{2g}{\Gamma + 2i\Delta_\text{mf}(\vec{r})}\mathcal{E}(\vec{r})
\end{align}
with the meanfield shift $\Delta_\text{mf}(\vec{r}) = \int d \vec{r}^{\prime} V(|\vec{r}-\vec{r}^{\prime}|)\rho(\vec{r}^{\prime})$.

\section{Absorption} \label{sec:absorption}
The propagation of the probe field amplitude $\mathcal{E}(r)$ close to an exciton resonance described by $\hat{X}$ is determined by 
\begin{align} \label{eq:motion_general1}
\partial_{t} \mathcal{E}(\vec{r})+\frac{c}{\bar{n}}\partial_z\mathcal{E}(\vec{r})&=-i\frac{g}{\bar{n}^2}\hat{X}(\vec{r})
\end{align}
with the refractive index $\bar{n}=2.74$ and the speed of light $c$. Taking expectation values, a closed absorption equation is obtained upon substitution of Eq.~(\ref{eq:exp_X}) into Eq.~(\ref{eq:motion_general1})
\begin{align} \label{eq:prop_E}
 \partial_z\mathcal{E}(\vec{r})&= -\frac{2g^2}{c\bar{n}\Gamma} \left( 1 - i \int d \vec{r}^{\prime} \rho(\vec{r}') \frac{V(|\vec{r}-\vec{r}^{\prime}|)}{\Gamma/2 + iV(|\vec{r}-\vec{r}^\prime|)} \right)\mathcal{E}(\vec{r}).
\end{align}
The absorption coefficient associated with the probe field intensity, $I(\vec{r}) = \hbar \omega_\text{in}c/\bar{n} |\mathcal{E}(\vec{r})|^2$, is then given by
\begin{align} \label{eq:alpha_full}
 \alpha &= \alpha_0 \left( 1 - \int d \vec{r}^{\prime} \rho(\vec{r}') \frac{(V(|\vec{r}-\vec{r}^{\prime}|)-2\Delta)V(|\vec{r}-\vec{r}^{\prime}|)}{\gamma^2/4+\Delta^2} g^{(2)}(\vec{r},\vec{r}^\prime) \right).
\end{align}
The first term describes the standard Lorentzian shape of an exciton resonance in the absence of a pump laser
\begin{align} \label{eq:alpha_0}
 \alpha_0 = \frac{4g^2 \gamma}{c\bar{n}(\gamma^2 + 4 \Delta^2)}.
\end{align}
The second term captures corrections to linear order in the density of pump excitons. We assume for simplicity that this density is constant across the sample, $\rho(\vec{r}') = \rho$. The integral boundaries can then be shifted to give the particularly simple form
\begin{align}
 \alpha = \alpha_0 \left( 1 - \rho \int \frac{(V(r)-2\Delta)V(r)}{\gamma^2/4+\Delta^2}g^{(2)}(r) d \vec{r} \right),
\end{align}
as given in Eq.~(2) of the main text. 

The corresponding meanfield absorption is obtained from the coherence given in Eq.~(\ref{eq:coherence_mf}) 
\begin{align}
 \alpha_\text{mf} = \frac{4g^2 \gamma}{c\bar{n}(\gamma^2+4(\Delta-\Delta_\text{mf})^2)}.
\end{align}
In effect, the meanfield approximation shifts the resonance position by $\Delta_\text{mf}$ with respect to the linear response, as can be seen by comparison with Eq.~(\ref{eq:alpha_0}), and as is illustrated in Fig.~1e of the main text.

\section{Universal absorption shape and scaling of $\beta$}\label{sec:beta}
The transmitted intensity after propagation through the crystal of length $L$ is given by Beer-Lambert's law $I=I_0 e^{-\alpha L}$. The pump-induced transmission difference is, therefore, directly obtainable from Eq.~(\ref{eq:alpha_full})
\begin{align}
 \Delta I = I_0 \left( e^{-\alpha L} - e^{-\alpha_0 L} \right) = I_0 e^{-\alpha_0 L} \left( e^{ \alpha_0\rho L \int \frac{(V(r)-2\Delta)V(r)}{\gamma^2/4+\Delta^2}g^{(2)}(r) d \vec{r}} - 1\right).
\end{align}
In the limit of low densities, the exponential in brackets can be expanded giving an expression for
\begin{equation}
\begin{aligned}
 \beta = \frac{\Delta I}{P} \approx I_0 e^{-\alpha_0 L} \alpha_0 L \frac{\rho}{P}\int \frac{(V(r)-2\Delta)V(r)}{\gamma^2/4+(V(r)-\Delta)^2} d \vec{r}.
\end{aligned}
\end{equation}
For exciton-exciton interaction of van der Waals type, $V(r)=C_6/r^6$, the integral can be simplified into
\begin{align} \label{eq:beta_scaled}
 \beta = I_0e^{-\alpha_0 L} 4\pi \alpha_0 L \frac{\rho}{P}r_\text{bl}^3 \int_0^\infty \frac{(1/\bar{r}^{6}-4\bar{\Delta})1/\bar{r}^{6}}{1+(1/\bar{r}^{6}-2\bar{\Delta})^2} \bar{r}^2 d \bar{r},
\end{align}
where we introduced dimensionless variables $\bar{\Delta} = \Delta/\gamma$ and $r = r_\text{bl} \bar{r}$ with the blockade radius $r_\text{bl}=\sqrt[6]{\frac{C_6}{\gamma/2}}$. This shows that the transmission difference $\Delta I$, or $\beta$, have a universal shape as a function of $\bar{\Delta}$, independent of the strength of the interaction. In particular, this proves the fixed position of the root and maximum across all principal quantum numbers as discussed in the main text and displayed in Fig.~1g therein.
We note that the linear absorption factor, represented by the exponential in Eq.~(\ref{eq:beta_scaled}), technically also influences the curve shape. However, since our main interest is in the pump-induced transmission maximum close around $\bar{\Delta} = 0$ and the root (not changed at all), this effect is small and is neglected in the following.

Eq.~(\ref{eq:beta_scaled}) also isolates the dependence on the principal quantum number $n$. In the limit of radiative scaling, optical dipole coupling scales as $g \sim n^{-\frac{3}{2}}$ and the linewidth scales as $\gamma \sim n^{-3}$, leaving a constant linear absorption $\alpha_0$. Thus, the $n$ dependence of $\beta$ for a given $n'$ mirrors that of the blockade radius
\begin{align}\label{eq:beta}
 \beta(n) \sim r^3_\text{bl}=\sqrt{\frac{C_6}{\gamma/2}}.
\end{align}
The van der Waals coefficient's dependence on probe and pump principal quantum number can be estimated from second-order perturbation theory \cite{Walther2018} from the probe [pump] exciton's dipole moment $d_\text{pr}(n) \sim n^{2}$ [$d_\text{pu}(n^\prime)\sim (n^\prime)^{2}$] and the pair state F\"orster energy $\delta(n,n^\prime) \sim n^{-3}+(n^\prime)^{-3}$ as
\begin{align}\label{eq:C6}
 C_6 (n,n^\prime) \sim \frac{d_\text{pr}^2 d_\text{pu}^2}{\delta} \sim \begin{cases}
n^7 (n^\prime)^4 \text{ for } n \ll n^\prime \\
n^4 (n^\prime)^7 \text{ for } n \gg n^\prime 
\end{cases}.
\end{align}
For deviations of $g^2$ and $\gamma$ from ideal scalings, as known~\cite{Kazimierczuk2014} from absorption lines in Cu$_2$O, one also has to consider the $n$ dependence of $\alpha_0$ in Eq.~\eqref{eq:beta_scaled}. 

\section{Asymmetry from phonon background}
The absorption lines of the yellow series in Cu$_2$O show an asymmetric curve shape due to an interference with a spectrally broad phonon background \cite{schone_phonon-assisted_2017}. This affects the shape of the pump-induced difference absorption as we show below. 
In the excitation process, an incident photon can either couple to the phonon background, described by a set of operators $\hat{X}_\vec{k}$, directly or first to a Rydberg exciton and then to the phonon continuum, thus creating a Fano resonance \cite{fano_effects_1961}.
We capture this by expanding Eq.~(\ref{eq:motion_mod}) and adding a phonon equation of motion
\begin{align}
 \partial_{t} \hat{X}(\vec{r}) &= -\frac{\Gamma}{2}\hat{X}(\vec{r}) - i g\mathcal{E}(\vec{r}) -i\sum_\vec{k}h_\vec{k}\hat{X}_\vec{k} - i \int d \vec{r}^{\prime} V(|\vec{r}-\vec{r}^{\prime}|)\hat{Y}^{\dagger}(\vec{r}^{\prime})\hat{Y}(\vec{r}^{\prime})\hat{X}(\vec{r}) \label{eq:eq_motion_fano1}\\
 \partial_t \hat{X}_\vec{k}(\vec{r}) &= -\frac{\Gamma_\vec{k}}{2}\hat{X}_\vec{k}(\vec{r}) - i g_\vec{k}\mathcal{E}(\vec{r}) -i h_\vec{k}\hat{X}(\vec{r}),
\end{align}
where $g_\vec{k}$ denotes the optical coupling rate to the phonons, $h_\vec{k}$ the exciton-phonon coupling and $\Gamma_\vec{k} = \gamma_\vec{k} - 2i\Delta_\vec{k}$ the complex phonon linewidth with linewidth $\gamma_\vec{k}$ and detuning $\Delta_\vec{k} = \omega_\text{in}-\omega_\vec{k}$. The light propagation is modified into
\begin{align} \label{eq:light_phonons}
 \partial_{t} \mathcal{E}(\vec{r})+\frac{c}{\bar{n}}\partial_z\mathcal{E}(\vec{r})&=-i\frac{g}{\bar{n}^2}\hat{X}(\vec{r}) - i \sum_\vec{k} \frac{g_\vec{k}}{\bar{n}^2}\hat{X}_\vec{k}(\vec{r}).
\end{align}
We start by considering the non-interacting system, $V=0$. 
In the steady-state, we can solve for the continuum operators
\begin{align}\label{eq:continuum_operators}
 \langle \hat{X}_\mathbf{k}(\mathbf{r}) \rangle = \frac{2}{\Gamma_\mathbf{k}} \left[ -ig_\mathbf{k} \mathcal{E}(\mathbf{r}) - i h_\mathbf{k} \langle \hat{X}(\mathbf{r})\rangle \right].
\end{align}
Across a single exciton resonance, the phonon states can for simplicity be assumed as flat ($g_\vec{k} = \text{const.}$ and $h_\vec{k} = \text{const.}$) and dense, allowing the sums to be approximated
by integrals and evaluated using Dirac's identity
\begin{align}
 \lim_{\epsilon \searrow 0} \int f(x) \frac{1}{x-i\epsilon} = i \pi \int f(x) \delta (x) + \mathcal{P} \int    \frac{f(x)}{x}.
\end{align}
In the limit $\gamma_\mathbf{k} \rightarrow 0$, this renders 
\begin{align}
 \sum_\mathbf{k} h_\mathbf{k} \langle \hat{X}_\mathbf{k}(\mathbf{r}) \rangle = i \pi \bar{h} \bar{g} \mathcal{E}(\mathbf{r}) + i \pi \bar{h}^2 \langle \hat{X}(\mathbf{r}) \rangle,
\end{align}
where we defined the coupling density $\bar{h} = h/\sqrt{|\delta \Delta_\vec{k}|}$ and, similarly, $\bar{g}$.
The polarization can be solved from Eq.~(\ref{eq:eq_motion_fano1})
\begin{align}
 \langle \hat{X}(\mathbf{r}) \rangle &= \frac{-ig + \pi \bar{h} \bar{g}}{- \frac{\Gamma}{2} + \pi \bar{h}^2} \mathcal{E}(\mathbf{r}),
\end{align}
giving rise to an asymmetric linear absorption spectrum
\begin{align}
 \alpha_\text{asym} = \frac{2}{c\bar{n}}\frac{(g^2-(\pi \bar{g}\bar{h})^2) (\gamma/2-\pi\bar{h}^2) - 2\pi g\bar{g}\bar{h}\Delta}{(\pi\bar{h}^2-\gamma/2)^2 + \Delta^2} + \frac{2}{c\bar{n}} \pi \bar{g}^2.
\end{align}
To simplify the expression we introduce $\bar{\gamma}/2 = \gamma/2 - \pi \bar{h}^2$, the asymmetry parameter $Q = -\pi \frac{\bar{g}\bar{h}}{g}$ as well as the constant background absorption coefficient $\alpha_\text{bg} = \frac{2}{c\bar{n}} \pi \bar{g}^2$ and exploit that the excitation rate is larger than the indirect rate $g \ll \pi \bar{g} \bar{h}$ 
\begin{align}
 \alpha_\text{asym} = \frac{4g^2}{c\bar{n}}\frac{\bar{\gamma} + 4Q\Delta}{\bar{\gamma}^2 + 4\Delta^2} + \alpha_\text{bg}.
\end{align}
We recognize a constant absorption term from the background as well as the plain exciton absorption from Eq.~(\ref{eq:alpha_0}). The cross term originates from the interference between direct and indirect excitation of the background and produces an asymmetric lineshape.
This shape corresponds to previous descriptions of the asymmetry in Cu$_2$O~\cite{toyozawa_interband_1964, jolk_linear_1998, ueno_contour_1969}. The linear spectra are used to extract $Q$ from experiment.

\begin{figure}
\begin{center}
 \includegraphics[width=\textwidth]{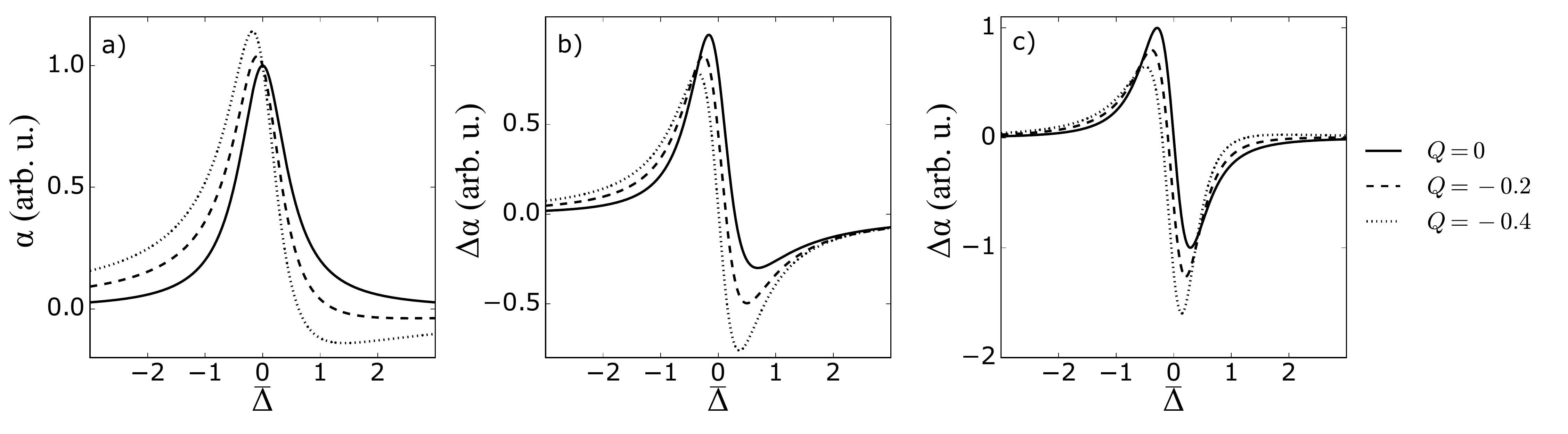}
\end{center}
\caption{{Fano asymmetry}: a) Linear absorption spectrum as a function of $\bar{\Delta} = \Delta/\bar{\gamma}$ for three different asymmetry parameters $Q$. The curves are normalized to the maximum of the Lorentzian profile ($Q=0$). 
	b) Normalized pump-induced differential absorption as a function of the asymmetry parameter $Q$ for van der Waals interactions ($\sim r^{-6}$), curves matching panel a). Maximum and root positions are red-shifted with increasing asymmetry but their difference decreases.
	c) Same as in b) but for direct dipole-dipole interactions ($\sim r^{-3}$).}
\label{fig:illustration_fano}
\end{figure}

The interacting system can be solved straightforwardly, following the approach outlined in Supplementary Notes \ref{sec:ex_corr}-\ref{sec:absorption} and giving the polarization
\begin{align} \label{eq:correlator_final}
 \langle \hat{X}(\mathbf{r}) \rangle &= \frac{-2ig(1 - i Q)}{- \bar{\gamma} +2i\Delta} \left[ 1 -i \int d \mathbf{r}^\prime \rho(\mathbf{r}^\prime) \frac{V(|\vec{r}-\vec{r}^\prime|)}{\frac{\bar{\gamma}}{2}-i\Delta+iV(|\vec{r}-\vec{r}^\prime|)} \right] \mathcal{E}(\mathbf{r}).
\end{align}
The absorption coefficient, here for flat densities, is readily obtained by inserting Eq.~(\ref{eq:correlator_final}) and Eq.~(\ref{eq:continuum_operators}) into Eq.~(\ref{eq:light_phonons})
\begin{align} \label{eq:asymmetry_final}
\alpha = \alpha_\text{asym} + \frac{2}{c\bar{n}} \Re \left[ \frac{2ig^2\left(1 - iQ \right)^2}{- \bar{\gamma} + 2i\Delta} \rho \int   \frac{V(r)}{\frac{\bar{\gamma}}{2}-i\Delta+iV(r)} d \vec{r} \right].
\end{align}
A non-vanishing asymmetry parameter, $Q\neq 0$, thus clearly leads to quantitative changes of the absorption features (Supplementary Fig.~\ref{fig:illustration_fano}). In particular, the maximum and root positions depend on $Q$. For van der Waals interactions, the difference between maximum and root decreases with $Q$. For direct dipole-dipole interactions of the form $V(r) = C_3/r^3$, this difference remains almost constant with $Q$. The corresponding predictions based on the experimental values of $Q$ are shown in Fig.~3 of the main text. We remark that the integral in Eq.~(\ref{eq:asymmetry_final}) for dipole-dipole interactions has a logarithmic divergence for large $r$, as can be seen by introducing a sufficiently large $R_1$ 
\begin{align}
 \int   \frac{V(r)}{\frac{\bar{\gamma}}{2}-i\Delta+iV(r)} d \vec{r} \approx 4 \pi \left[ \int_0^{R_1} \frac{V(r) r^2}{\frac{\bar{\gamma}}{2}-i\Delta+iV(r)} dr 
 + \lim_{R_2 \rightarrow \infty} \frac{C_3}{\frac{\bar{\gamma}}{2}-i\Delta} \ln \left( \frac{R_2}{R_1} \right) \right].
\end{align}
However, the nonlinearity is dominated by the second term whose shape as a function of $\Delta$ is determined by
\begin{align}
 \Re \left[ -i\frac{\left(1 - iQ \right)^2}{(\frac{\bar{\gamma}}{2} -i\Delta)^2} \right].
\end{align}

\clearpage

\section{Relation to exciton-in-plasma studies}
\begin{figure}
	\begin{center}
		\includegraphics[width=\textwidth]{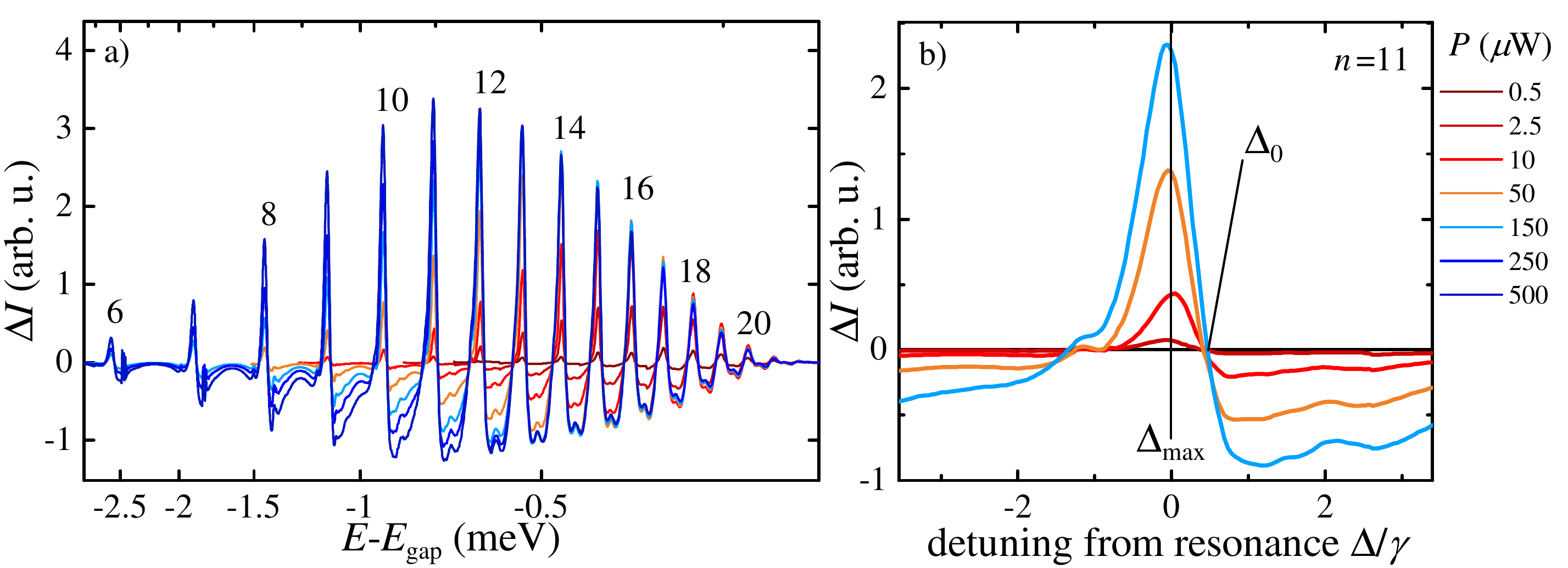}
		\includegraphics[width=0.8\textwidth]{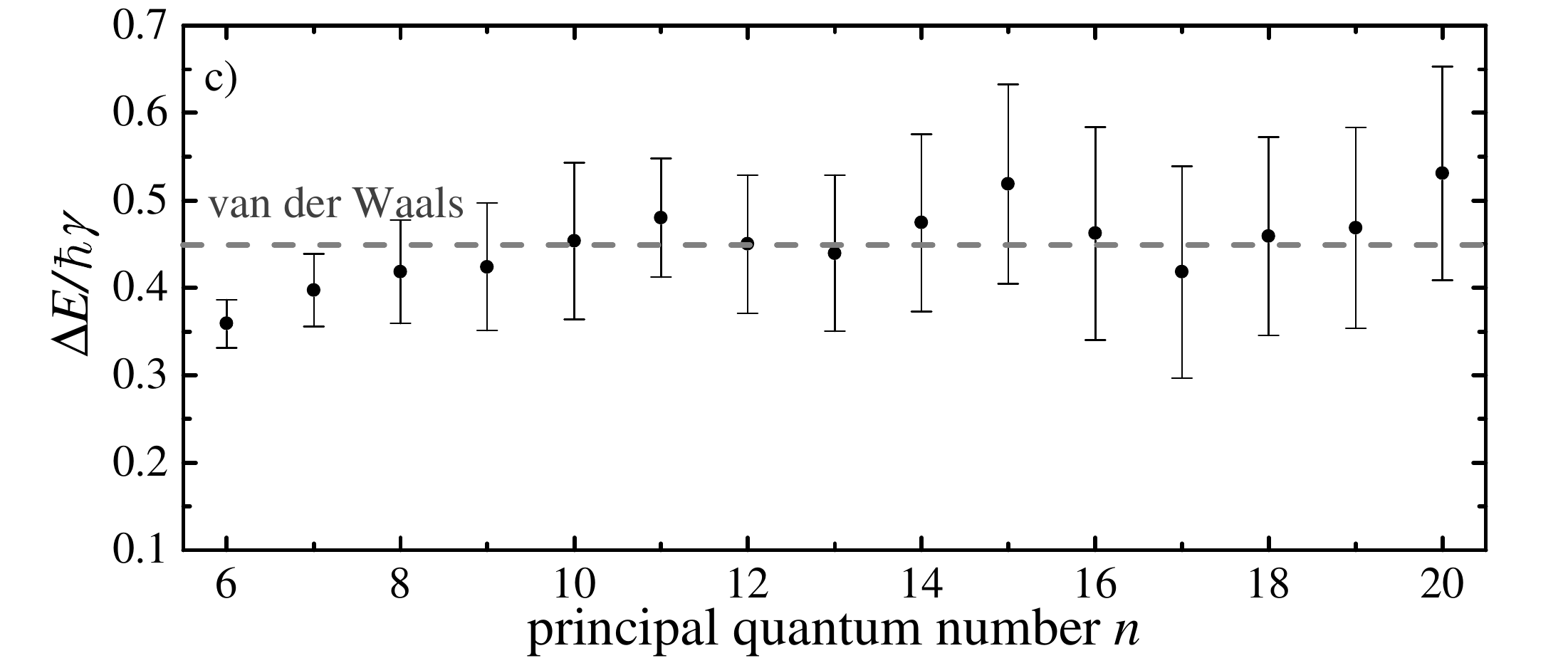}
	\end{center}
	\caption{{Differential signal (plasma)}:
		a) Differential transmission $\Delta I$ for excitons $n=6$ up to $n=21$, recorded with a pump laser energy fixed above the band gap. The pump power $P$ is increased from 0.5 to 500~$\mu$W. 
		b) Closeup of the $n=11$ resonance shown for pump powers in the range of linear power dependence. 
		Also here, the curve shape is universal and exhibits a fixed extreme point at $\Delta_\text{max}$ and a universal root on the high-energy side at $\Delta_0$, where the probe absorption is independent of the pump intensity.
		c) The experimentally obtained ratio $\Delta E/\hbar\gamma$ is comparable to the theoretical value of $0.45$ (grey dashed line), that is predicted for the van der Waals potential. }
	\label{fig:SI4}
\end{figure}

Strong optical excitation of the crystal by a pump laser beam above the band gap unavoidably leads to the creation of a density of free electron-hole pairs that may additionally influence the exciton states as studied in Ref.~\cite{Heckoetter2018}. Compared to Ref.~\cite{Heckoetter2018}, we restrict our analysis to very low pump powers, where the differential transmission spectra are linear in the applied pump power.
In order to estimate the impact of such an electron-hole plasma on the observed spectra we show a second set of differential transmission spectra recorded with a pump laser energy at 2.20 eV, i.e., 28 meV above the band gap. 
We repeat the same analysis as in the case of resonant pumping of Rydberg excitons (main text). 
Supplementary Fig.~\ref{fig:SI4}a) shows the recorded spectra for pump powers from 0.5 to 500~$\mu$W. 

First, we investigate the spectral features of the recorded resonances. At first glance, there are no major differences compared to the pump scenario shown in Fig.1d of the main text. 
We again find a non-shifting maximum growing with increasing pump power and a fundamental root at $\Delta_0$ on the high-energy side, as shown exemplarily for the $n=11$ resonance in Supplementary Fig.~\ref{fig:SI4}b). 
These characteristic signatures result solely from power-law interaction potentials and thus stand in contradiction to a possible dynamical screening induced by a plasma, 
which can be described to first approximation by a Debye potential~\cite{kremp_quantum_2005} $V(r)\sim e^{-\kappa r}/r$. 
Further, we also determine the universal quantity $\Delta E/\hbar\gamma$ for each resonance and find values that are comparable to the expectations given by the van der Waals interaction of 0.45, as shown in Supplementary Fig.~\ref{fig:SI4}c). These observations cannot be explained by dynamical screening. 

\begin{figure}
	\begin{center}
		\includegraphics[width=\textwidth]{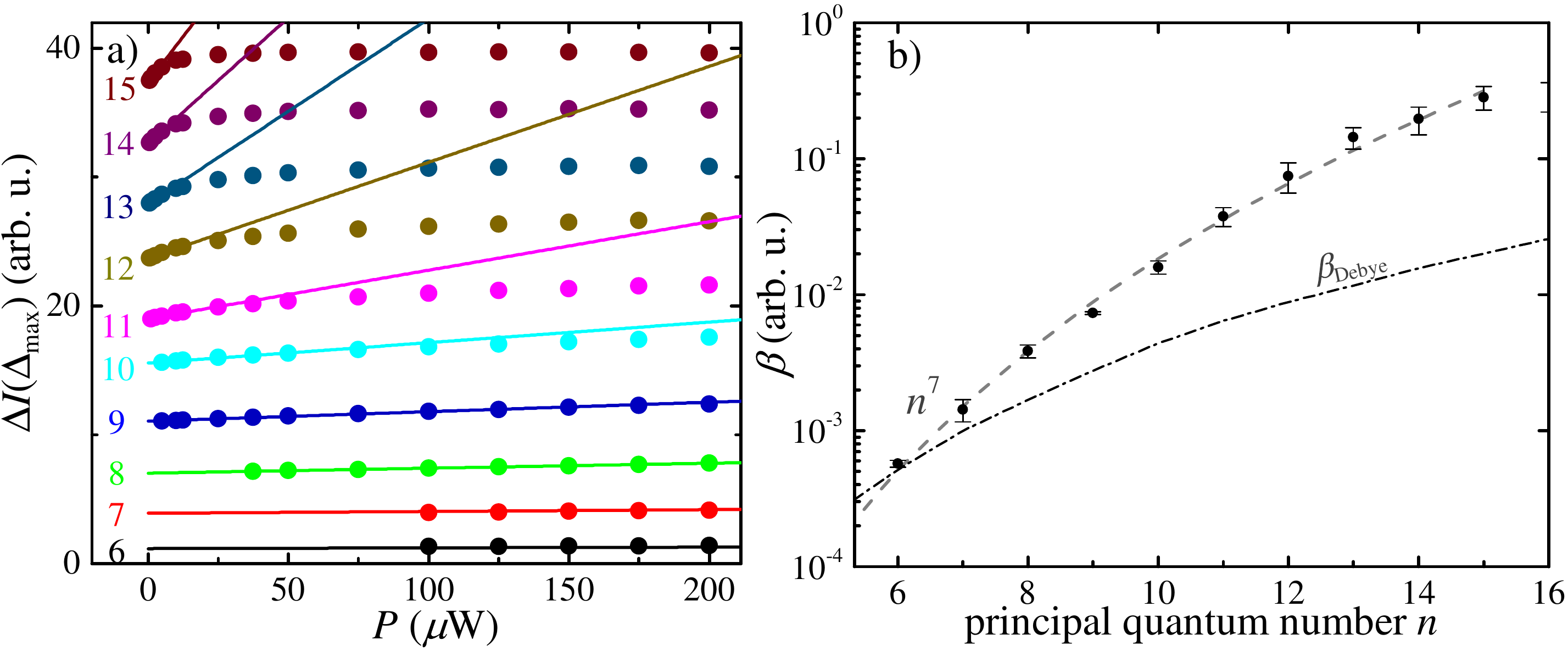}
	\end{center}
	\caption{{Characteristic scaling (plasma)}:
	a) Maximum differential signal $\Delta I(\Delta_\text{max})$ as a function of pump power $P$ for principal quantum numbers $n=6$ to 15. The pump laser energy is set above the band gap. 
	The solid lines show fits with a slope $\beta$ in the range of a linear dependence on pump power for each resonance.
	b) Experimental scaling of $\beta$ as a function of principal quantum number $n$. 
	The observed scaling follows an $n^7$ dependence in the range from $n=6$ to 15 as indicated by the grey dashed line. 
	The black dashed-dotted line shows the scaling of $\beta_\text{Debye}$ predicted by the Debye model. 
	It is much weaker than observed in the experiment. 
	The curves are shifted by an arbitrary value to coincide at $n=6$ for comparison. 
}
	\label{fig:SI5}
\end{figure}

Following the analysis in the main text, we now evaluate the power dependence of the maxima of each resonance in order to determine a characteristic scaling law of the slope $\beta$. 
For the states from $n=6$ to $n=15$, we find a pump power regime of linearly increasing maxima, whose range decreases drastically with $n$ (Supplementary Fig.~\ref{fig:SI5}a)). 
Remarkably, the experimentally determined linear slope $\beta$ shows a strong increase as a function of principal quantum number $n$ scaling approximately as $n^7$ for low and intermediate states (Supplementary Fig.~\ref{fig:SI5}b), grey dashed line). 
From the Debye model, we numerically extract a pump power dependence of the matrix elements $g^2 \approx g^2_0 + g^2_1 P$ with $g^2_1 \sim n^{0.85}$. This implies a scaling of $\beta \propto e^{-\alpha_0 L} \frac{g^2_1}{\gamma}$ to lowest order in $P$.
The radiative expectation on the scaling of $\beta \sim n^{3.85}$ is corrected by the experimental values of both $\alpha_0$ and $\gamma$ in Supplementary Fig.~\ref{fig:SI5}b), demonstrating that the Debye model significantly underestimates the observed slope of $\beta$.  

Interestingly, the observed scaling is even slightly steeper than the one expected for van der Waals interactions with resonantly excited Rydberg pump excitons at fixed $n'$. In this case the scaling for low $n$ is $\beta(n,n)\propto n^5$, cf. Eqs.~\eqref{eq:beta} and \eqref{eq:C6}. 
The stronger optical response observed in the experiment may be explained by fast relaxation of free charges into an unknown distribution of excitons with different principal quantum numbers and angular momenta \cite{Takahata2018, killian_formation_2001}. 

In conclusion, also for above-bandgap excitation we find characteristic spectral signatures and an $n$-dependent scaling of the signal strength that are in accordance with van der Waals-type interaction between excitons created by fast relaxation of free electron-hole pairs.

\clearpage

\bibliographystyle{naturemag}
\bibliography{RydbergBib}